# The European Union, China, and the United States in the Top-1% and Top-10% Layers of Most-Frequently-Cited Publications: Competition and Collaborations

Loet Leydesdorff,[1]* Caroline S. Wagner,[2] and Lutz Bornmann [3]

**Abstract**

The percentages of shares of world publications of the European Union and its member states, China, and the United States have been represented differently as a result of using different databases. An analytical variant of the Web-of-Science (of Thomson Reuters) enables us to study the dynamics in the world publication system in terms of the field-normalized top-1% and top-10% most-frequently-cited publications. Comparing the EU28, USA, and China at the global level shows a top-level dynamics that is different from the analysis in terms of shares of publications: the United States remains far more productive in the top-1% of all papers; China drops out of the competition for elite status; and the EU28 increased its share among the top-cited papers from 2000-2010. Some of the EU28 member states overtook the U.S. during this decade; but a clear divide remains between EU15 (Western Europe) and the Accession Countries. Network analysis shows that internationally co-authored top-1% publications perform far above expectation and also above top-10% ones. In 2005, China was embedded in this top-layer of internationally co-authored publications. These publications often involve more than a single European nation.

**Keywords**: world share, citation analysis, global science, excellence, measurement, country, Europe, USA, China

[1] Amsterdam School of Communication Research (ASCoR), University of Amsterdam, Kloveniersburgwal 48, 1012 CX Amsterdam (The Netherlands); loet@leydesdorff.net ; *corresponding author.
[2] Battelle Center for Science & Technology Policy, John Glenn School of Public Affairs, Ohio State University, Columbus, OH (USA); wagner.911@osu.edu
[3] Division for Science and Innovation Studies, Administrative Headquarters of the Max Planck Society, Hofgartenstr. 8, 80539 Munich (Germany); bornmann@gv.mpg.de

## 1. Introduction

In 2011, the Royal Society—the national science academy of the UK—using data from the Scopus database (an Elsevier product) issued a report showing that China was on a trend line to overtake the USA in terms of numbers of publications by 2013 (Clarke *et al.*, 2011; Plume, 2011). Along the same lines, Hill *et al.* (2007) and Wagner (2011) showed that several European nations had increased their overall citation shares, and six European countries had overtaken the USA in terms of relative citation rates. In this study, we explore these trends further by examining patterns among the most-highly-cited papers, expecting to find that country shares among the most elite papers reflect specific historical patterns (Bornmann, de Moya Anegón, and Leydesdorff, 2010). Moreover, we expect to find that the most elite scientists are highly networked internationally (Wagner, 2008).

This shift away from U.S. dominance within global publication and citation shares has attracted scholarly attention. The discussion has focused on the drop in the share percentage of the USA in scientific databases (Shelton & Foland, 2009; Leydesdorff & Wagner, 2009) and the exponential growth of Chinese contributions (Moed, 2002; Zhou & Leydesdorff, 2006). Using Web of Science (WoS, a Thomson-Reuters product) data, Leydesdorff (2012) argued that the growth of China may have been overestimated by the Royal Society (cf. Moed *et al.*, 2011). An extrapolation by Shelton and Leydesdorff (2012) suggested a date beyond 2020 for the cross-over, and noted that the exponential growth of Chinese scientific publications had slowed to linear growth rates during the 2000s. The shares attributed to the European Union are less clear, partly because the borders of the EU continue to change given the accession of ten new member states in 2004, Romania and Bulgaria in 2007, and a further expansion to the EU28 most recently with the accession of Croatia in 2013.

The two multidisciplinary indexing services—Scopus and WoS—have kept pace with the growth of global science by expanding their coverage. In 2009, for example, WoS announced a regional expansion to cover more journals from Central and Eastern European countries (Testa, 2011) in response to increased coverage by Scopus, that itself was launched only in 2004. The addition of new journals was also an attempt to address the English language bias in the database (Van Leeuwen *et al.*, 2000).

A number of studies have discussed the influence of international collaborations on the global system of science (Adams, 2013; Glänzel, 2001; Luukkonen *et al.*, 1993; Okubo *et al.*, 1992; Wagner & Leydesdorff, 2005). Multiple national addresses may partly account for the changes in shares attributed to countries (Persson *et al.*, 2004); but the growth of the databases and different (e.g., fractional) counting methods alone cannot account for the changes in relative positions among nations. The increased effects of networking in terms of co-authorship relations among member states of the EU has also been used as an indicator of further integration at the European level (Frenken, 2002; Frenken & Leydesdorff, 2004; Hoekman *et al.*, 2010).

In this study, we address the question of whether the shifting patterns hold also for the top-1% and top-10% segments of the most-highly-cited publications (that is, articles, reviews, and letters). These top-segments of the publication and citation curves represent the scientific elite, which some have argued functions as a special structure, citing one another differently from lower strata in the publication system (Cole, 1970; Mulkay, 1976). According to the results of Bornmann, de Moya-Anegón, and Leydesdorff (2010), highly-cited work in all scientific fields tends to cite highly-cited papers more than medium-cited work. In this study, we explore how some leading nations participate in this "elite" structure of most-highly-cited publications, and whether and how this structure is influenced by and/or overlaps with international collaborations.

The *Science and Engineering Indicators* report issued by the U.S. National Science Board provides percentages of the top-1% most-highly-cited publications for 2002 and 2012 (National Science Board, 2014: Appendix table 5-57) and the numbers of publications for 13 broad fields of science and engineering in terms of six percentile rank classes (top-1%, top-5%, top-10%, top-25%, top-50%, and bottom-50% in Appendix table 5-58; cf. Bornmann & Mutz, 2010). We take a similar approach to compare the national addresses of papers in the top-1 and top-10 percentile rank classes, but add the dynamic perspective of a decade of years (2000-2012) and include international co-authorship relations in the evaluation.

The data used in this study were harvested from an analytical version of WoS developed and maintained by the Max Planck Digital Library (MPDL, Munich). This database includes the Science Citation Index-Expanded (SCI-E), the Social Sciences Citation Index (SSCI), and the Arts and Humanities Citation Index (A&HCI) of Thomson Reuters since 1980. However, the citation impact of all papers is "field"-normalized against reference sets using the 226 WoS Categories (WC) that are attributed by Thomson Reuters to the 10,000+ journals in WoS (cf. Leydesdorff & Opthof, 2011).[1] The percentile values can be compared because they are normalized for differences among fields of science, document types, and citation windows. This organization of the data allows us to construct timelines of field-normalized impact scores (e.g., top-10%) for different nations, for groups of nations (such as the EU), and for international collaborations.

---

[1] On March 3, 2014, the SCI covered 8,623 journals, the SSCI 3,134 journals, and the Arts & Humanities Citation Index 1,727 journals. The overlap between the SCI and the SSCI is on the order of 600 journal titles.

We explore the longitudinal development of the comparison between the EU28, USA, and China at the global level, and of the decomposition of the EU28 both in terms of member states and as a network of international co-authorship relations. We thus add the perspectives of using the proportions of top-1% and top-10% publications ($PP_{top-1\%}$ and $PP_{top-10\%}$; Waltman *et al.*, 2012; cf. Tijssen *et al.*, 2002) to the analysis in terms of percentages of world shares of publications. We did not include other nations (e.g., Japan and South Korea) in the discussion. In a follow-up study, we repeated this analysis with a focus on the BRIC(S) countries (Brazil, Russia, India, China, and South Africa; Bornmann, Wagner, & Leydesdorff, in press).

## 2. Methods and materials

We used integer counting to allocate publications to a country whenever this country's name is present in the publication's address lines.[2] Integer counting allows us to assume that *ceteris paribus*, 10% of a nation's internationally co-authored publications can be expected to belong to the top-10% most-highly-cited set (and *mutatis mutandis* for the top-1% percentile class).[3] One limitation of integer counting is that the percentages cannot be expected to add up to one hundred because internationally co-authored publications are attributed to more than a single country (Anderson *et al.*, 1988); the disadvantage of fractional counting, conversely, can be a negative effect of internationalization on the performance thus measured, *ceteris paribus* (Leydesdorff, 1988).

---

[2] In the case of fractional counting, each country receives a fractional count based upon the number of country names in the address lines (e.g., Anderson *et al.*, 1988; Braun *et al.*, 1989; cf. Irvine *et al.*, 1985). For example, if a record contains three addresses of which two are in the USA and one is in China, the record is attributed for 2/3$^{rd}$ to the USA and 1/3$^{rd}$ to China. In version 5 of WoS (since 2011), it is possible to fractionate in terms of the number of authors, since this version contains information to relate sequential authors unambiguously to address information (cf. Costas & Iribarren-Maestro, 2007).
[3] In the case of fractional counting, the contributions of nations other than the ones under study have to be taken into account for the specification of an expectation.

Citation distributions are extremely skewed and therefore non-normal (Seglen, 1992). The use of percentile classes has been proposed as a non-parametric alternative to normalization on the basis of arithmetic averages of citation counts (Bornmann & Mutz, 2011; Pudovkin & Garfield, 2009; Zitt *et al.*, 2005). We delineate reference sets for calculating the percentile rank classes in terms of the (226) WoS categories of journals, data types (articles, reviews, or letters), and publication years. Percentile values are calculated using Hazen's (1914:1550) method (Bornmann, Leydesdorff, & Mutz, 2013). Field-normalized indicators can be used for comparisons over time (Schubert and Braun, 1986).

Note that the percentile values are paper-specific. The normalized citation impact for a set of papers requires the specification of appropriate reference sets for each paper. When papers in the dataset are published in journals assigned to more than a single WoS category (and thus more than one percentile value would be available), the average of the percentile values is used (by MPDL). As noted, we focus on the proportions of papers in the 90$^{th}$ percentile rank class (PP$_{top10\%}$) and 99$^{th}$ (PP$_{top1\%}$). However, PP$_{top-1\%}$ and PP$_{top-10\%}$ values are provided only until 2010 in order to include at least three years (2011-2013) in the citation window (Wang, 2013).

The data for the years 2000-2012 were downloaded from the database between January 15 and February 15, 2014. Within the EU28 set, double counts because of international collaboration were corrected, and the UK contribution was corrected for collaborations based on English, Scottish, Welsh, or Northern-Irish addresses. Co-authorship networks are visualized using Pajek.[4] The network analysis is static since it is limited to 2005 as a single year.

The observed value of $PP_{top-10\%}$ can be compared with an expected value of 10% in each set (and similarly 1% for $PP_{top-1\%}$). In principle, the observed and expected values can be tested against each other for statistical significance using the *z*-test for independent proportions (Leydesdorff & Bornmann, 2012; cf. Sheshkin, 2011: 656). However, the numbers of publications for countries are usually so large that most values will be statistically significant; significance is then no longer informative (Schneider, 2013).

We use the expected values of 1% and 10%, respectively, as a benchmark for the interpretation of the observed $PP_{top-1\%}$ and $PP_{top-10\%}$. For example, an observed $PP_{top-10\%}$ of 15% is 5 percentage points above expectation; the observed *versus* expected ratio is then 15/10 or 1.5. One can use this observed/expected ratio as a summary measure of evaluation since values above and below unity are above or below expectation, respectively. Of course, many countries in the world cannot be expected to have publications in either the top-10% or top-1%.

## 3. Results

We proceed in three steps:

---

[4] Pajek is a program for network analysis and visualization freely available for non-commercial use at http://pajek.imfm.si/doku.php?id=download .

1. We first compare the USA, China, and the EU28 in terms of their percentages of shares of world publications (articles, reviews, and letters) in WoS in both the top-layers of 1% and 10%. We show that the dynamics among these three geographical regions are very different when measured in terms of $PP_{top-1\%}$, $PP_{top-10\%}$, or percentages: The main competition is now between the USA and the EU28, and not with China.

2. The EU28 is thereafter disaggregated in terms of its member states. Whereas the larger member states such as the UK, France, and Germany show a pattern of decline in terms of percentages of shares as expected (given the increased competition), some Mediterranean countries have improved their shares in terms of publications during the 2000s. In terms of highly-cited publications, European nations are improving their shares in the top-10% more than in the top-1%, but some smaller countries are improving the top-1% share in their portfolios above expectation. Accession countries do not (yet?) participate in this growth in the top layers.

3. The network analysis of international co-authorships among EU member states, China, and the USA shows first that the network *structures* in the top-10% and top-1% layers are similar to the network based on all publications: the USA can be considered as a major partner of all EU nations, and China is also well connected, albeit most frequently to the USA. However, the observed numbers of co-authored publications among most nations are above expectation (in 79.1% of the cases). Furthermore, the observed/expected ratios at the 1% level are much higher than at the 10% level, supporting the thesis that this top-1% layer can be considered as an elite structure with a dynamic different from those of less-cited collaborations (Bornmann *et al.*, 2011; Wagner, 2008). Co-authorship relations between China and these western nations (EU member states and the USA) also score above expectation in terms of both $PP_{top-1\%}$ and $PP_{top-10\%}$ in a large majority of the networked relations.

### 3.1 Comparison: EU28, China, and USA

Let us first turn to the global comparison among the aggregated EU28, China, and the USA both in terms of shares of publications and the shares in $P_{top-1\%}$ and $P_{top-10\%}$. Figures 1a and b show the data using percentages of shares of all publications between 2000 and 2012 (Figure 1a) *versus* the top-1% and top-10% most-frequently-cited ones (in Figure 1b). In both figures, the growth of China is approximately similar (0.8 percentage points growth per year) and linear ($R^2 \approx .99$). The yearly growth in China's top-1% contribution is slightly higher: 0.85 percentage points per year. Figure 1a shows the well-known trend towards a cross-over between the USA and China, but this is foreseen beyond 2020 using this data. As noted above, the growth of the Chinese share of global science is no longer exponential, but closely resembles a linear approximation.

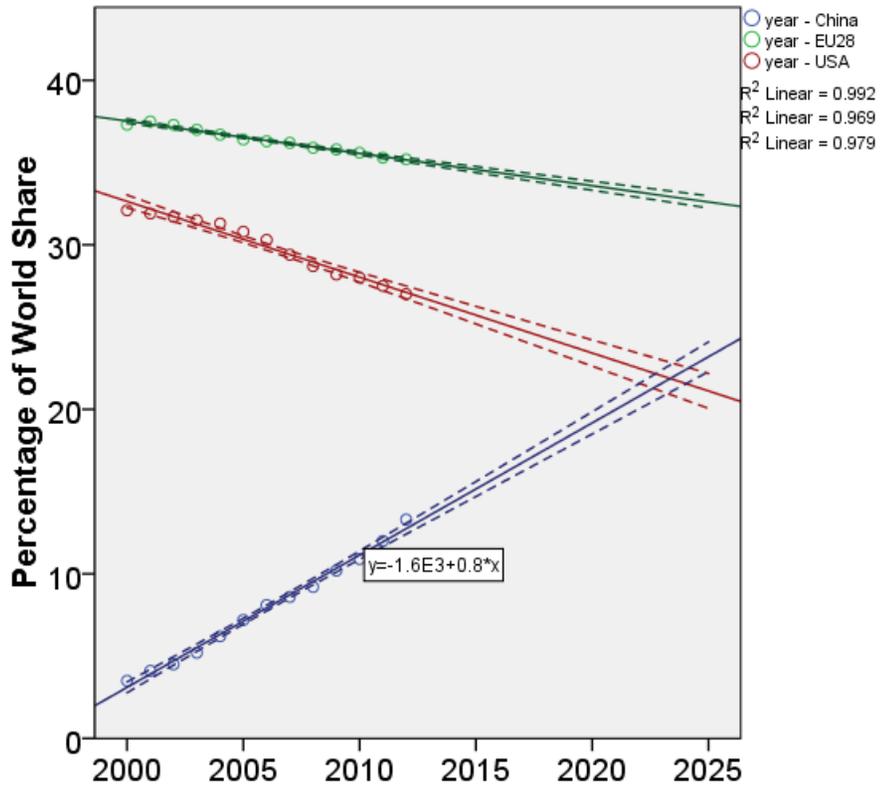
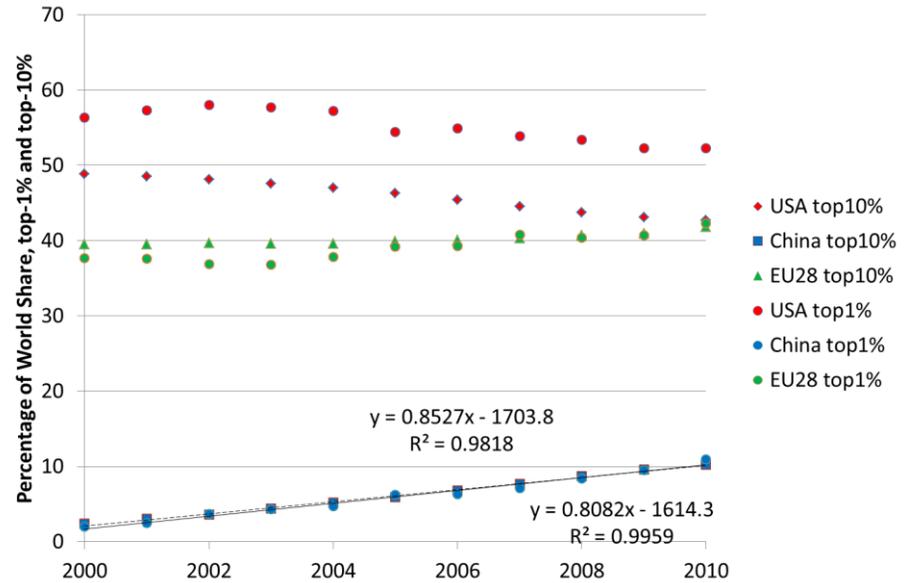

**Figures 1a and b**: Percentages of world share of publications from China (blue), USA (red), and EU28 (green; in Figure 1a on the left), and in $PP_{top-1\%}$ and $PP_{top-10\%}$ (Figure 1b to the right).

The impressive growth of China (and other emerging countries) in global science depresses the percentage shares of both the EU28 and USA in Figure 1a because this is a zero-sum game (despite the integer counting) where some lose as others gain. However, Figure 1b shows a different dynamic: the EU28 gains gradually in the top-10% segment at the expense of the USA. One can expect a cross-over between the EU28 and the USA in the near future within the top-10% segment. However, the distance between the U.S. and the EU is much larger in the top-1% segment, with the U.S. dominating. In these top segments, the competition between China and the USA or Europe is not an issue (Leydesdorff & Wagner, 2009).

We will decompose the EU28 into its member states in a later section, and then raise the question of whether certain member states carry the EU into the higher impact classes more than others. But let us first look at $PP_{top\text{-}1\%}$ and $PP_{top\text{-}10\%}$ relative to expected values (in Figure 2). Assuming that 10% of a country's papers would be part of the top-10% for stochastic reasons, Figure 2 shows the observed values in relation to this benchmark on the left axis, and against the expectation of 1% for the top-1% along the right axis.

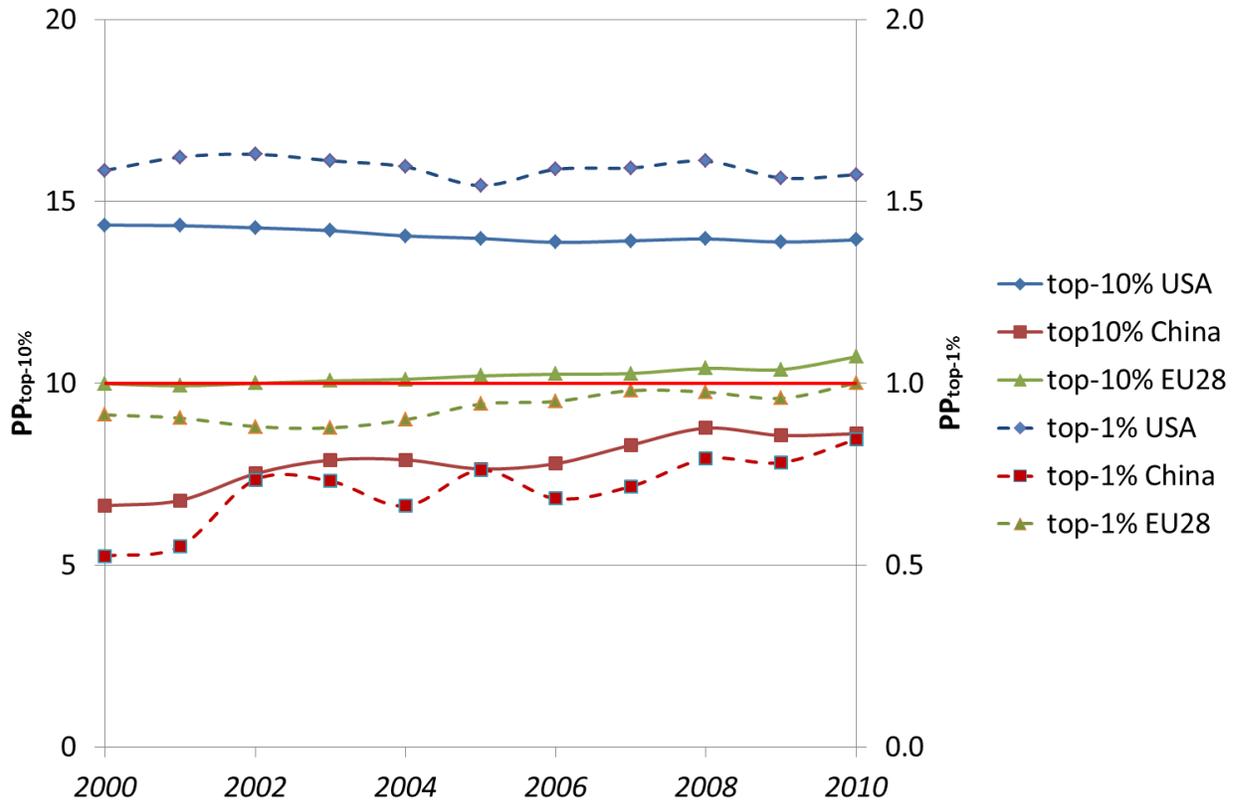

**Figure 2**: PP$_{top-10\%}$ (solid lines, left axis) and PP$_{top-1\%}$ (dashed lines, right axis) 2000-2010 for the USA, China, and EU28 compared with the expectation (red horizontal line).

Figure 2 shows that China has been able to *improve* its proportions in the top segments. The figure also shows that the USA is overwhelmingly dominant from this perspective. Despite their improved position in recent years, China's share of top-cited papers is still below the worldwide average and also below the EU28's average performance; the EU28 performs very close to the expectation. Unlike the data represented in Figure 1, the percentages of Figure 2 do not a represent a zero-sum game because the comparison is now not in terms of percentages of world shares of publications, but we assume a percentage of each country's total numbers of publications.

Only in the case of the USA are the two curves for $PP_{top-1\%}$ and $PP_{top-10\%}$ meaningfully different: in this case the top-1% rises far more above expectation than seen in the top-10%, whereas for both the EU28 and China, the top-1% sets are more or less a proportional subset of the top-10%. In other words, the Chinese and European shares in the high-impact classes $PP_{top-1\%}$ and $PP_{top-10\%}$ are embedded in the variation; but in the case of the USA, the most-highly-cited publications exhibit a dynamics different from those of the other publications. The higher one comes in the citation ranks, the more likely the papers will have an American address.

*3.2 Decomposition of the EU28*

The European Commission stimulates projects among member states; and one would expect these R&D programs to lead to more international co-authorship relations in the resulting publications. Nevertheless, the member states have retained also a national character (Frenken & Leydesdorff, 2004; Glänzel, 2001). The national systems vary considerably in size, with the UK and Germany being major contributors to world science on one side of the distribution, and small nations such as Malta or Cyprus on the other extreme. We analyze data for all 28 member states in terms of $P_{top-1\%}$ and $P_{top-10\%}$, but limit the visualization and discussion first to the seven major players in Europe's science efforts: the UK, Germany, France, Italy, Spain, Netherlands, and Sweden. The cutoff in terms of size is a bit arbitrary—we could have added Poland, Belgium, or Denmark—but the scale for the other 21 smaller countries would have to be different in the visuals.

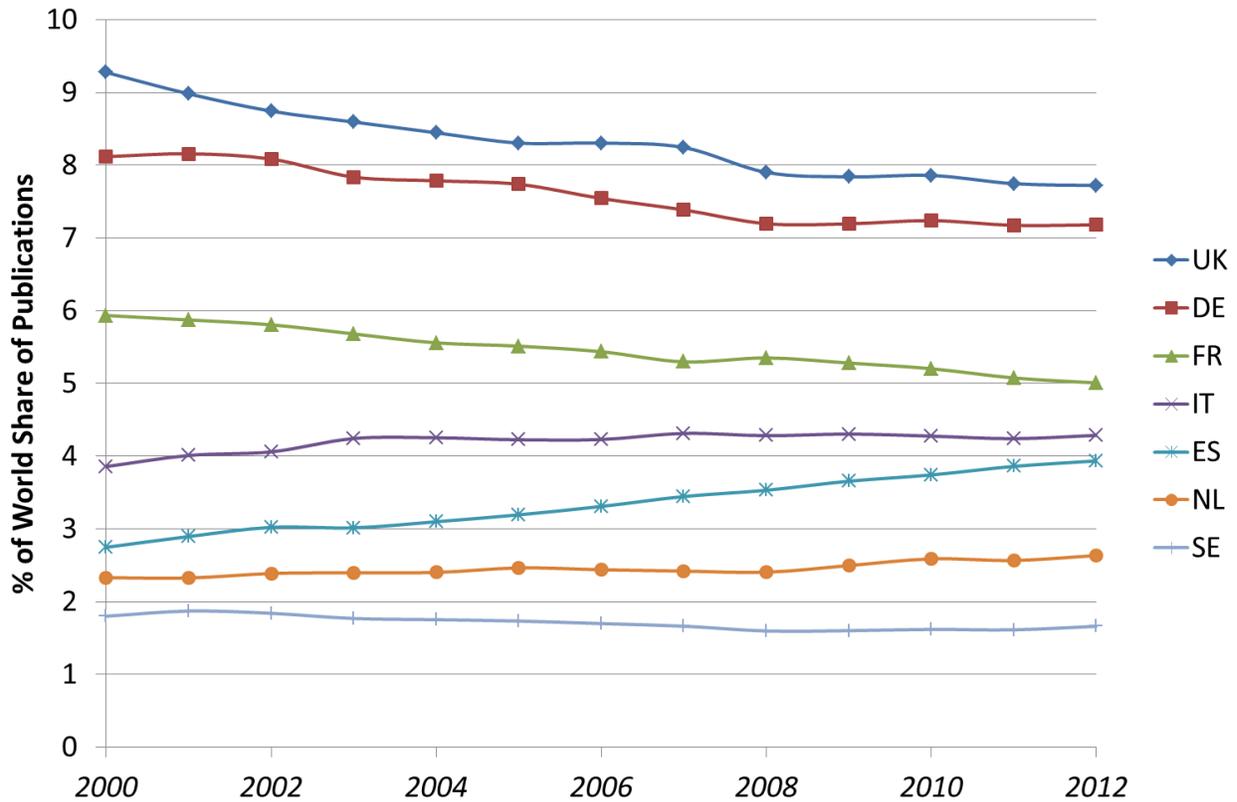

**Figure 3:** Percentages of world shares of some larger and medium-sized European nations.

Figure 3 shows the relative decline of the EU nations in terms of world share, with the notable exception of Spain.[8] The decline of the aggregated EU28 (visible in Figure 1a above) is mainly due to the downward slope shown for the major players, that is, the UK, Germany, and France, whose positions have been affected by the rapid increase of contributions from China and other emerging players. Italy and the Netherlands, however, have a marginally upward slope. (A downward trend can also be indicated for Japan, but is not shown here.)

---

[8] Similarly, not shown here, Portugal has been continuously improving its percentage of world share from 0.4% in 2000 to 0.9% in 2012.

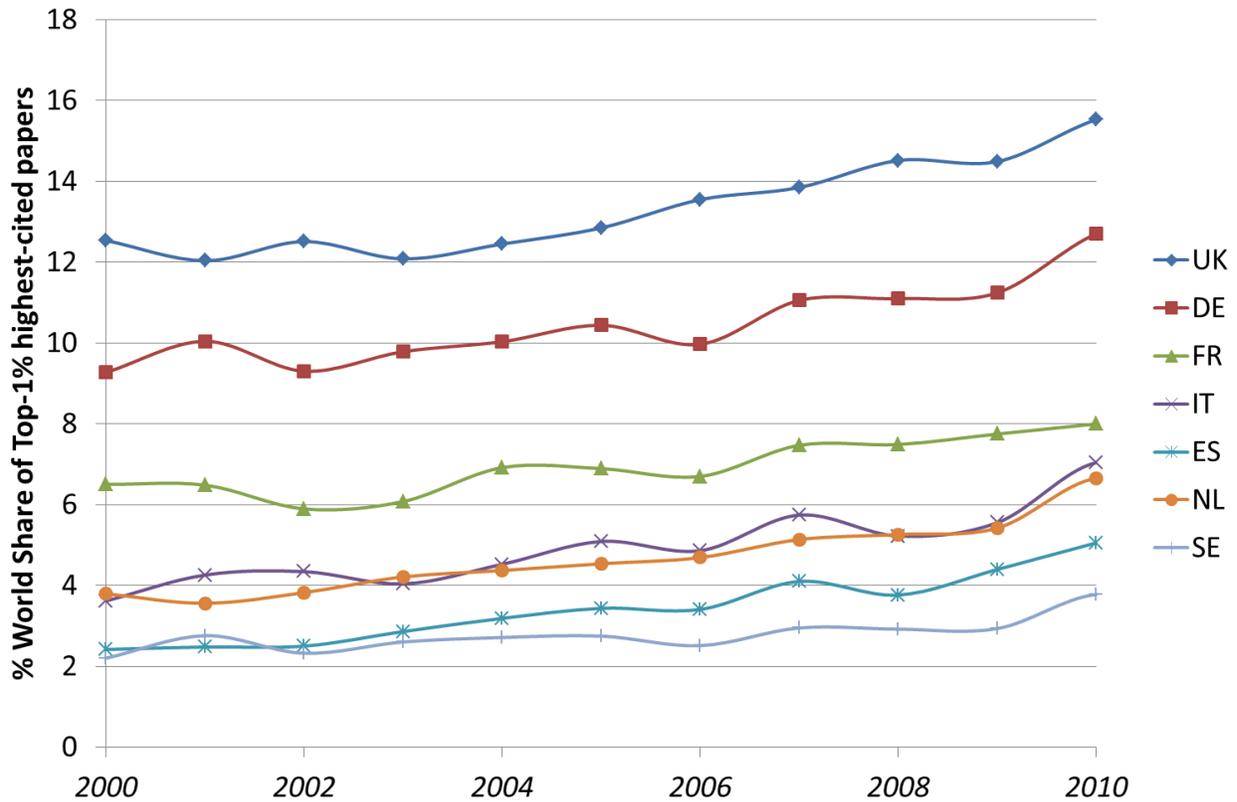

**Figure 4:** World share of the top-1% most-cited publications for some larger and medium-sized smaller European nations.

Figure 4 shows that the situation is very different in the PP$_{\text{top-1\%}}$: many European countries are able to improve their positions within this indicator. The effect is more pronounced in the top-1% than the top-10% (not shown here). In other words, a trend is visible among the European nations towards more-highly-cited publications. As we shall see below, this trend is not limited to these seven countries.

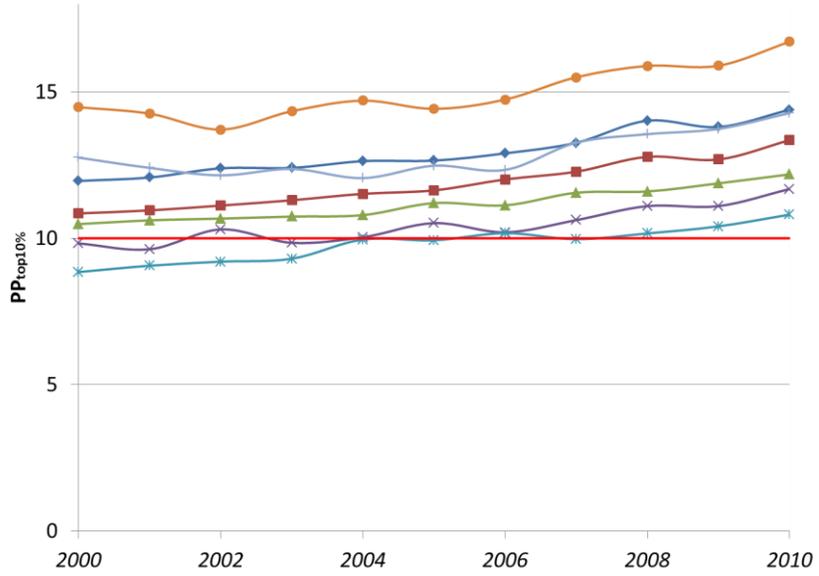 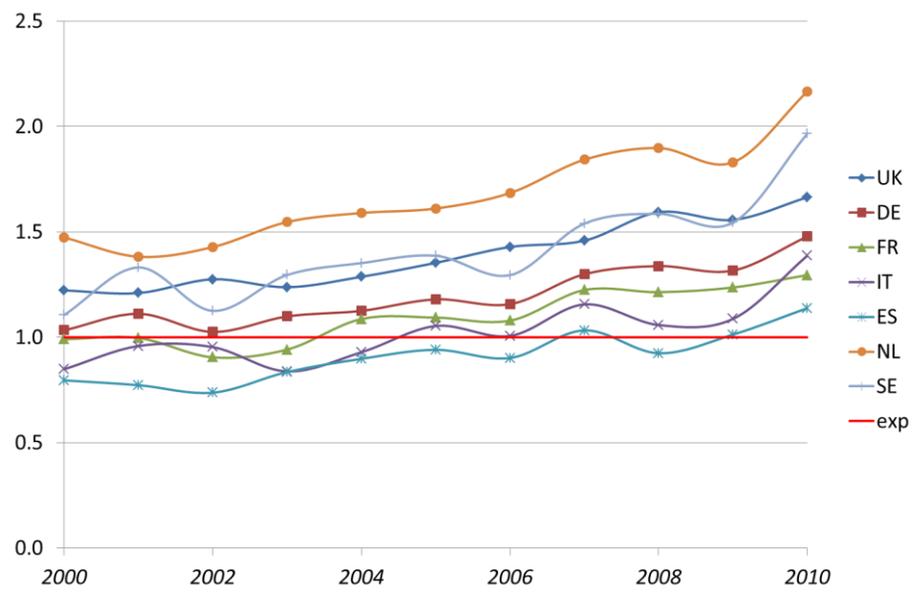

**Figures 5a and b**: Development of $PP_{top-10\%}$ (left) and $PP_{top-1\%}$ (right) for seven medium-sized and smaller EU nations. The expected values of 1% and 10% are marked by a red line.

Figures 5a and b show that increases in the high-quality segments were consistent among these seven member states over the timeframe studied. $PP_{top-1\%}$ increases faster, but less continuously than $PP_{top-10\%}$. (However, one should consider that $PP_{top-1\%}$ involves approximately ten times fewer papers than $PP_{top-10\%}$.) The Netherlands and Sweden perform in the top segments above expectation. The patterns for the UK, Germany, and France are similarly upward, but in this percentile rank class, Spain and Italy show this upward trend as well.

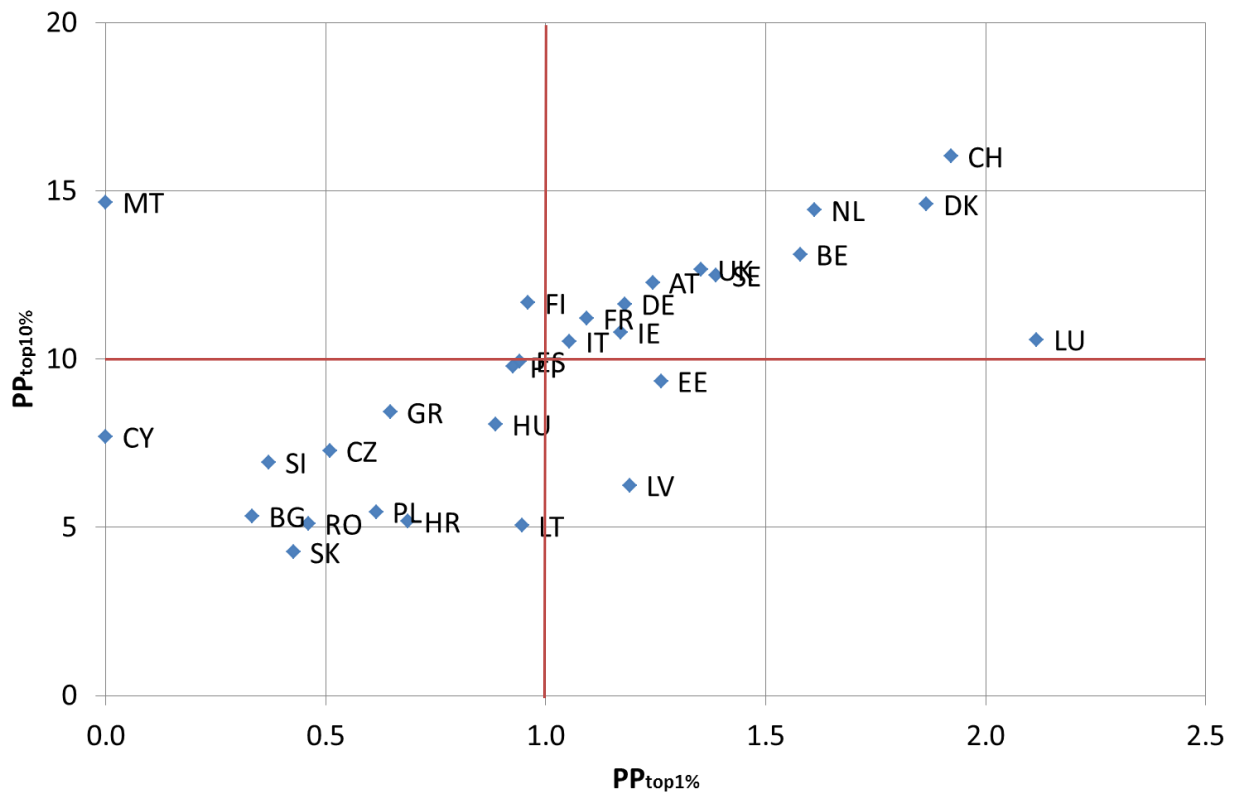

**Figure 6**: $PP_{top-10\%}$ plotted against $PP_{top-1\%}$ for 28 EU member states in 2005. We added Switzerland (CH) to this figure.

Figure 6 aggregates the contributions of $PP_{top-1\%}$ and $PP_{top-10\%}$ for the EU28 in 2005, relative to the benchmarks of 1% (on the horizontal axis) and 10% (on the vertical axis), respectively. *Grosso modo*, countries in Western Europe are indicated in the first quadrant (above expectation on both indicators) and Accession Countries in the third quadrant (below expectation). Italy, Spain, and Portugal are represented close to the origin. Finland manages to score above expectation within the top-10%, but not within the top-1%. Estonia and Latvia participate in the top-1% set above expectation.

Switzerland (CH) is added to this figure because May (1997) mentioned Switzerland as an outlier at the elite end. Indeed, this is confirmed by these results. Cyprus and Malta have no papers belonging to the top-1% in 2005. Denmark and Luxembourg were not included in Figures 4 and 5 above (because of the lower numbers), but are also highly represented in terms of $PP_{top1\%}$. In this representation, the USA would be positioned close to the Netherlands (NL) and China close to Hungary (HU). Let us now consider whether and how these two large nations (the USA and China) are woven into the European networks of co-authorship relations.

*3.3 Co-authorship networks within the EU28 in relation to China and the USA*

The 2005 co-authorship network among the member states of the EU28, China, and the USA is visualized in Figure 7. The sizes of the nodes are proportional to the total number of co-authorships for the respective countries.[1]

---

[1] The total number of co-authorships is computed by using the weighted degree centrality as a vector in Pajek. The weight is equal to the number of bi-lateral co-authorship relations represented by respective cell value in the co-authorship matrix.

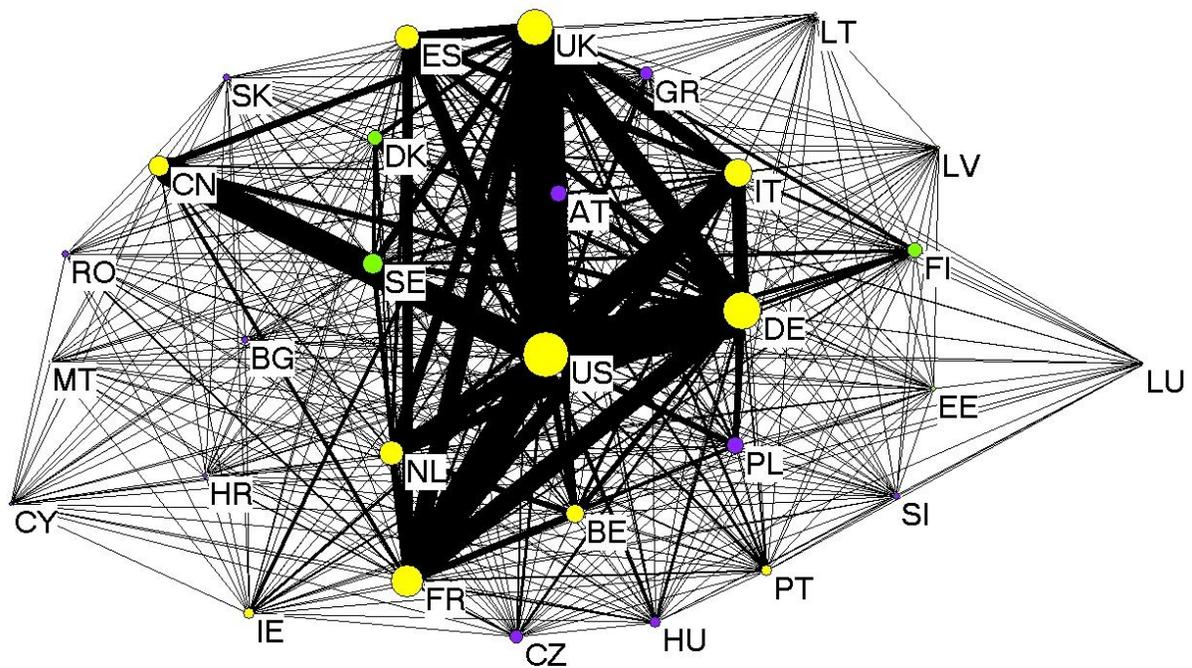

**Figure 7:** Co-authorship network in 2005 among EU28 countries, the USA, and China; $Q = 0.0253$ (*N of Clusters* = 3; Blondel *et al.*, 2008); node sizes proportional to the number of international co-authorship relations in the set; Kamada & Kawai's (1989) force-directed graph-drawing algorithm was used for the layout.

Figure 7 shows that the USA is tightly woven into the network of leading European nations in science, whereas China is integrated into the network through the USA, although China has relationships with all European nations except Luxembourg. (The relations with Latvia and Malta are based on single publications.) In other words, virtually all relations exist and the network is tightly-knit into a single (strong) component. The average degree—that is, the number of linkages—among these 30 countries is 27.7; the density of the matrix is accordingly high (0.92) and the clustering coefficient is 0.96.

Although the modularity of this highly-connected network is consequently low ($Q = 0.0253$), the community-finding algorithm of Blondel *et al.* (2008) enables us to distinguish three groups: (1) a central group (shown in yellow in Figure 7) with almost all the EU15 countries (in Western Europe), the USA, and China; (2) a group of Nordic countries (Scandinavia and the Baltic states); and (3) the Accession Countries. Austria is placed among the latter group in terms of international co-authorship relations in this (!) year.

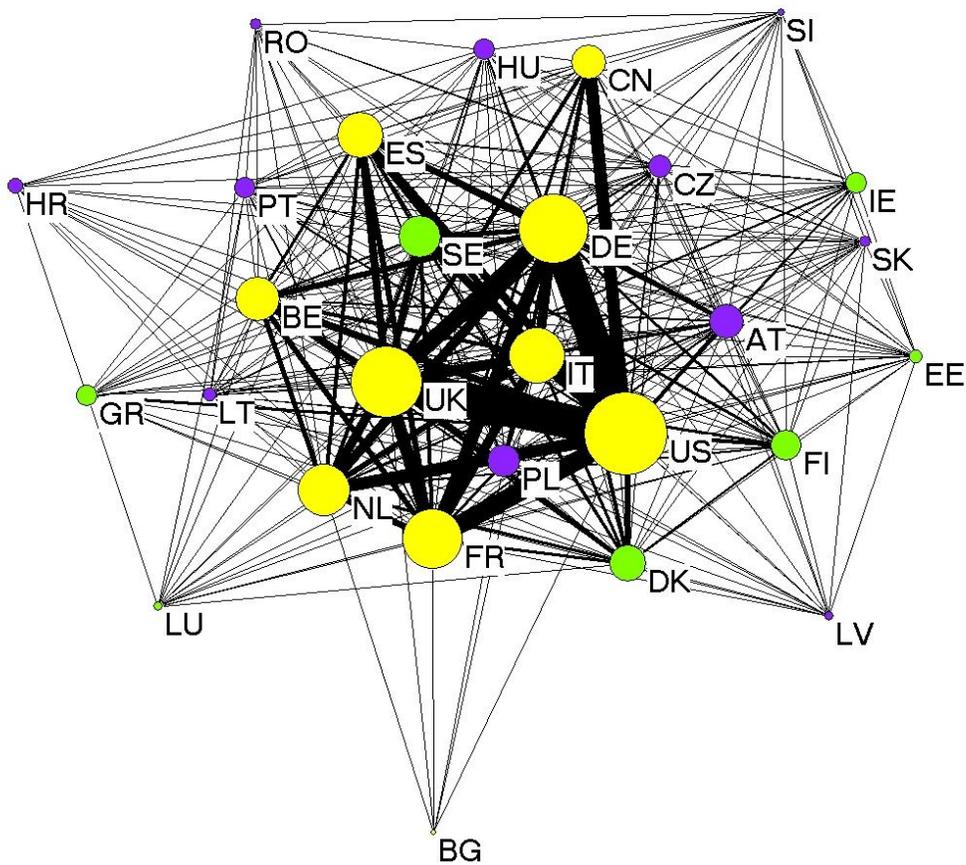

**Figure 8**: $P_{top1\%}$ of the international co-authorship network among EU member states, China, and USA; $Q = 0.02623$ (Blondel *et al.*, 2008); node sizes proportional to the number of international co-authorship relations in the set; Kamada & Kawai's (1989) force-directed graph-drawing algorithm was used for the layout.

Precisely the same classification is found in the top-10% layer of the network (not shown here), but the structure of the network is somewhat different in the top-1% (Figure 8). Malta and Cyprus are no longer found when the analysis is limited to this top-layer (because these nations do not have papers in the top-1% set in 2005; see Figure 6). Ireland, Greece, and Luxembourg are now placed in the second (previously Nordic) group, whereas Portugal is attributed to the third group (of mainly Accession Countries). Otherwise this network of the remaining 26 EU member states, China, and the USA is very similar in structure to the one shown in Figure 7.[2] The top-10% layer provides only an in-between configuration and therefore is not shown here.

Whereas the structure of the networks at the three levels (all papers, $PP_{top10\%}$, and $PP_{top1\%}$) is rather similar, the numbers in the top-1% and top-10% layers are far above expectation. In both cases, only 20.9% of the possible network values in the lower triangle of the co-authorship matrix (= n * (n-1)/2) is below expectation.[3] The mean of the observed/expected values is 1.83 (± 1.29) in the case of the top-10% values and even 5.24 (± 6.86) in the case of the top-1% values. The high standard deviation in the latter case indicates that the distribution is skewed, but the line for the top-1% level is above the one for the top-10% across almost the entire range in Figure 9. This means that the numbers of top-1% highly-cited papers are far more above expectation when compared with papers in the top-10% segment. In other words, this multi-national top-layer is indicated as an elite structure similar to the top-line for the USA in Figure 2.

---

[2] The density is 0.78; average degree is 22.0; and the clustering coefficient is 0.88.
[3] Malta and Cyprus are not present in the top-1% layer, and thus the number of countries involved is in this case 26 + 2 (China and the USA), and the number of observations is 28 * 27/2 = 378 (that is, the lower triangle of the matrix) against 30 * 29/2 = 435 for the top-10% layer.

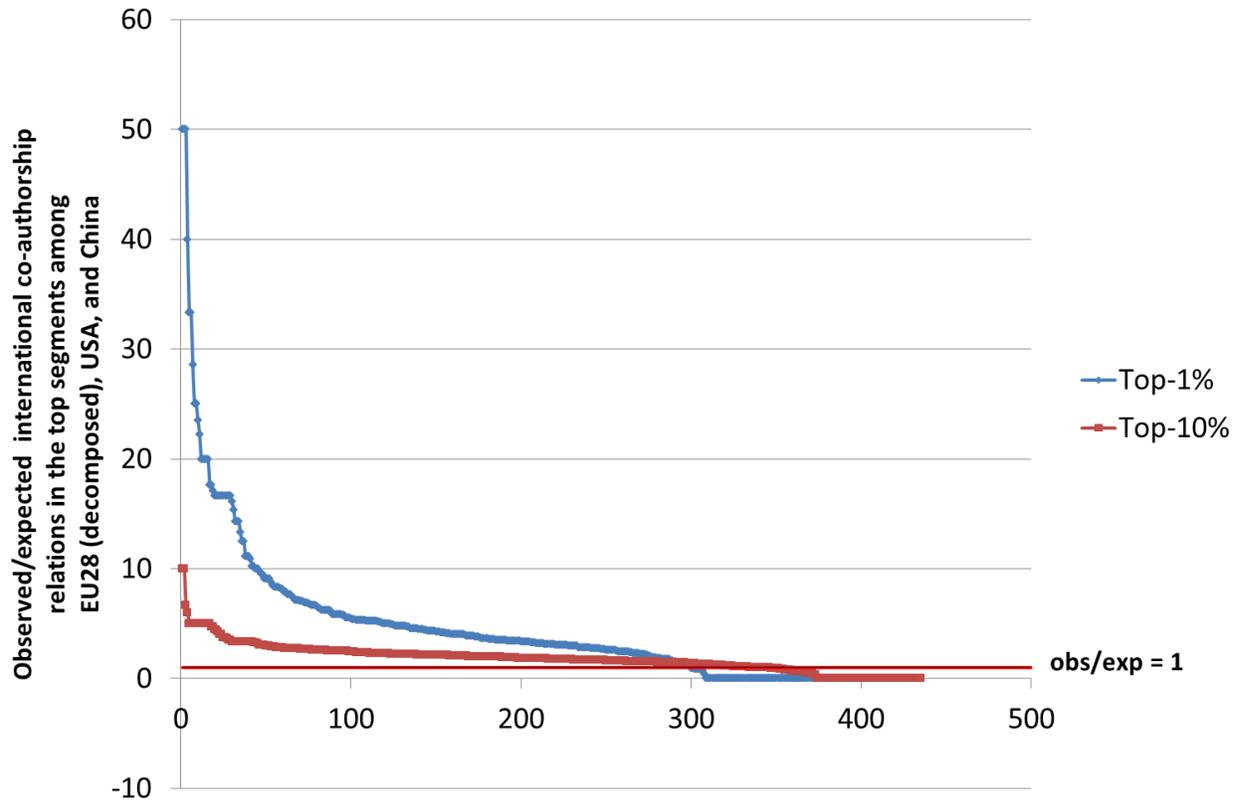

**Figure 9**: Observed/expected ratios of international co-authorship relations among 28 EU member states, USA, and China in 2005.

For example, Figure 9 shows a first outlier value of 50 for only two co-authorship relations between Luxembourg and Portugal, of which one belongs (perhaps by chance) to the top-1% of most-frequently cited papers. However, among the 430 publications co-authored in 2005, for example, between authors with an Italian and Danish address, 34 or 7.9% belong to the top-1% category; and 112 or 26.0% belong to the top-10% set against an expectation of (430/10 =) 43. In other words, sometimes the expectations are low numbers, but the enhancement of the top-layer in terms of international co-authorship relations is very significant across the set.

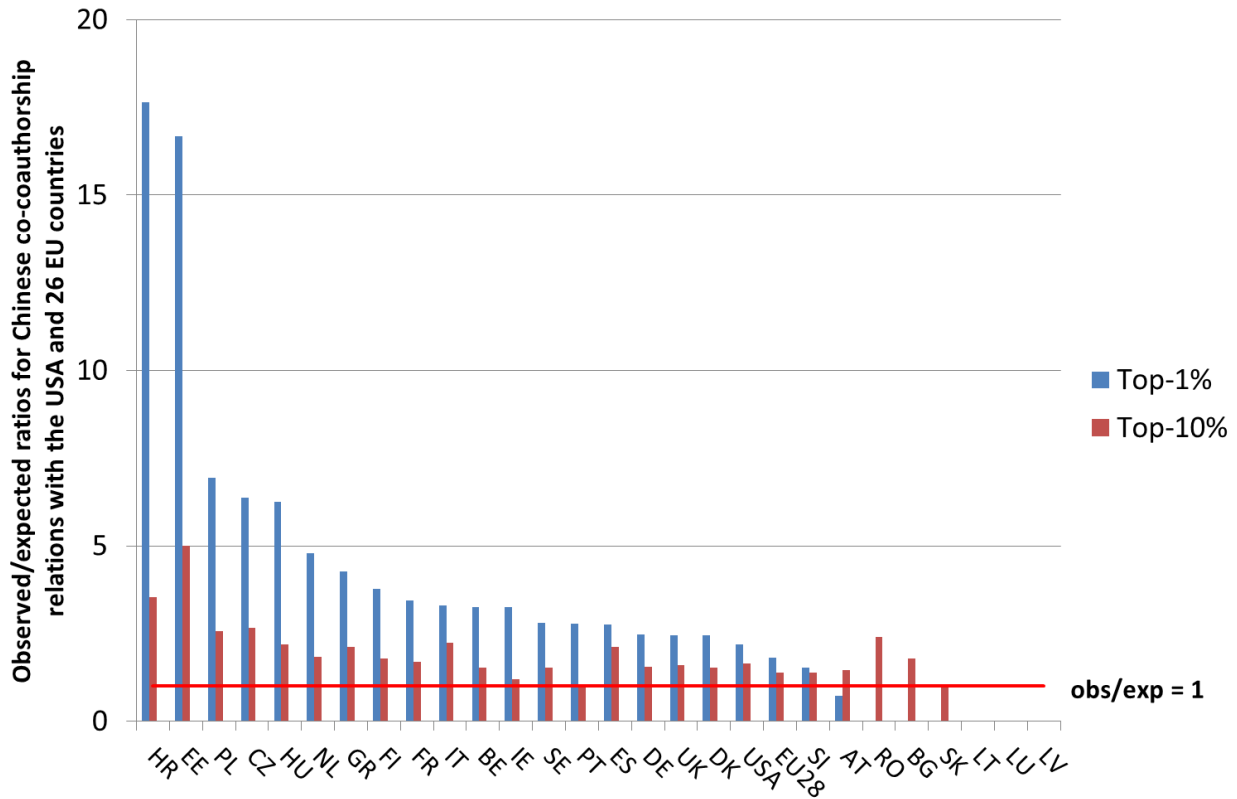

**Figure 10**: Observed/expected values for Chinese co-authorship relations with the USA and 26 EU member states in the top-1% and top-10% segments. (Malta and Cyprus not included.)

Chinese collaborations with co-authors from almost all EU countries and the United States can be found in the top-10% and top-1% segments as shown in Figure 10. Chinese participation in this top-layer falls almost always above expectation. Notably, the outliers are with small Accession Countries such as Croatia (HR) and Estonia (EE) because of the low numbers in the denominator (6 and 34, respectively). These large values at the left of Figure 10 are likely dominated by co-authorships with multiple EU nations, as well as relatively small numbers per country. However, 144 publications with a Chinese address are co-authored, for example, with an author from Poland, of which ten score in the top-1% layer (against an expectation of 1.44) and 37 in the top-10% layer (against an expectation of 14.4). China's international reach can be seen in this data as early as 2005.

Note that neither the USA nor the aggregated EU28 are performing much above expectation among the top co-authoring nations in Figure 10: Chinese-US co-authorship in the top layers is above expectation, but less so than with many European nations, including some Accession Countries. One can compare the co-authorship relations of China with the USA and EU28 by aggregating the data for the EU member states in the top-1% layer using the so-called "shrink network" option in Pajek. However, this alternative procedure does not correct for double counting generated by co-authorship relations with more than a single European member state (Table 1).

**Table 1**: Sum of the EU member states *versus* the aggregated EU28 (corrected for more than a single European address) in the top-1% segment in 2005.

|  | *Sum of EU member states* | *EU28 aggregated* |
|---|---|---|
| EU addresses | 3,149 | 1,047 |
| US addresses | 1,093 | 1,093 |
| Chinese addresses | 177 | 177 |
| Within-EU28 co-authorships | 3,533 | |
| EU-US co-authorships | 1,732 | 1,004 |
| EU-Chinese co-authorships | 203 | 88 |
| US-Chinese co-authorships | 134 | 134 |
| EU-US-CN "triadic" co-authorships* | 152** | 45 |

* Triadic relations were counted also in the numbers of the bilateral ones.
** Because of integer counting, a single paper can carry more than one triad with different EU nations.

The resulting figures show that Atlantic partnership in co-authored publications is far more important in the top-1% layer than collaborations with China. The collaborations between China and Europe in terms of bi-lateral relations among nation states (203) are considerably larger than those between China and the USA (134); but when the EU28 is aggregated, the US-Chinese relations are more important. The relation between the aggregated EU and the USA is balanced in terms of the number of papers involved (1,047 and 1,093, respectively).

## 5. Conclusions and discussion

We focused in this study on the top-1% and top-10% layers in terms of numbers of publications involving some of the major players in science. Because of the policy relevance in the discussions about competition among the USA, Europe, and China, we chose these as our geographical units and additionally decomposed the EU28 in terms of its member states. However, we did not include other nations (e.g., Japan and South Korea) in the study. In a follow-up study, we repeated the analysis with a focus on the BRIC(S) countries (Brazil, Russia, India, China, and South Africa; Bornmann, Wagner, & Leydesdorff, in press).

The analytical version of the WoS database allowed us to test and quantify our previous conjectures about the top-layers of scientific publications (Bornmann *et al.*, 2010; Wagner, 2008): the percentile values can be compared because they are normalized for differences among fields of science, document types, and citation windows. Our results suggest that $PP_{top-1\%}$ or $PP_{top-10\%}$ do not function as specific layers of publications with citation rates far above expectation in the case of China or EU28. However, they do so in the USA, and to a larger extent in the top-1% layer than in the top-10%. The citation dynamics therefore are very different in the USA than in China or the EU28 (at the aggregated level); but we saw that this is changing across Western Europe, and particularly in some of the smaller EU nations such as Denmark and the Netherlands. The dynamics in the USA is different because the system is differentiated in terms of an elite layer of publications, whereas in the other case the most-highly-cited publications are more embedded in the variation.

Disaggregation of the EU28 into its member states made it possible to show that a set of core countries in the EU15 leads the upward tendency of the EU28 in terms of $PP_{top-1\%}$ or $PP_{top-10\%}$. The Mediterranean countries (Spain, Italy, and Portugal) are performing almost as expected, while the Accession Countries and Greece remained below expectation (in Figure 6). The Baltic states, however, are deviant among the Accession Countries and performing notably better in terms of $PP_{top-1\%}$ perhaps because of international co-authorship relations in this segment with the other Nordic (that is, Scandinavian) countries.

In the network analysis of international co-authorship relations, this Nordic group was indicated as different from the core set of the EU15 using a community-finding algorithm (Blondel *et al.*, 2008; cf. Leydesdorff, Park, & Wagner, in press). China and the USA, however, could not be distinguished from the core group of EU15 nations either on the basis of all papers or at the levels of top-1% and top-10%. The Accession Countries form a third group; one or two of the EU15 countries are also part of this third grouping.

More importantly than these relatively minor changes in the tightly-knit network of co-authorship relations among these nations are the unexpectedly high observed-versus-expected values in the networks among top-1% and top-10% highly-cited papers. In some cases the high values are caused by low values in the denominator (that is, the expectation), but overall the pattern (in Figure 9) is even more pronounced than in the case of the USA as presented in Figure 2: papers in the top-1% layer are present far above expectation, and even more so than in the top-10% layer. This means that internationally co-authored papers among these countries are significantly more part of an elite structure among scientific publications even if they are from authorship relations with the Accession Countries or China (which are expected to remain otherwise below expectation in terms of citation rates).

This effect of international collaborations among these advanced nations (including nowadays China) was more or less intuited in the literature (Bornmann *et al.*, 2010; Wagner, 2008), but not to such an extent. Numerous studies (e.g., Narin *et al.*, 1991) showed that co-authored papers attract more citations than singled-authored papers and internationally co-authored papers even more so. A testable question for follow-up research would be whether the proportion of the internationally co-authored publications in the top-1% or top-10% publications is significantly higher than the proportion of nationally co-authored papers, which in turn is significantly higher than the proportion of single-authored papers.

In summary, the elite structure of top-cited publications is primarily American, but also some European nations can increasingly participate in it. The other nations participate in this top-layer with a higher probability through international collaborations, but to a lesser extent so when publishing without foreign co-authorship.